\documentclass[10.5pt,compsoc]{JSC}
\graphicspath{{figures/}}
\usepackage{graphicx}
\usepackage{footmisc}
\usepackage{subfigure}
\usepackage{url}
\usepackage{multirow}
\usepackage[noadjust]{cite}
\usepackage{amsmath,amsthm}
\usepackage{amssymb,amsfonts}
\usepackage{booktabs}
\usepackage{color}
\usepackage{ccaption}
\usepackage{booktabs}
\usepackage{float}
\usepackage{fancyhdr}
\usepackage{caption}
\usepackage{xcolor,stfloats}
\usepackage{comment}
\usepackage{cuted}  
\usepackage{captionhack}
\usepackage{epstopdf}
\usepackage{times}
\usepackage{enumerate}
\usepackage[inline]{enumitem}

\usepackage{amsmath}
\usepackage{enumerate}
\usepackage[inline]{enumitem}
\usepackage{graphicx}
\usepackage{subcaption}
\usepackage{algorithm2e}
\usepackage{float}
\usepackage{amsfonts}
\usepackage{amssymb}
\usepackage[utf8]{inputenc}
\usepackage{url} 
\usepackage{color, colortbl}	
\definecolor{LightCyan}{rgb}{0.88,1,1}
\usepackage{natbib}
\setcitestyle{numbers}
\setcitestyle{square}
\setcitestyle{comma}

\usepackage{blindtext} 
\headevenname{\zihao{-5}{\textbf{\emph{}}}}%
\headoddname{{\sf Narmadha Mohankumar et al.:}\quad {\textbf{\emph{Principal Component Regression to Study the Impact of Economic Factors on Disadvantaged Communities}}}}%

\newtheoremstyle{mystyle}{0pt}{0pt}{\normalfont}{1em}{\bf}{}{1em}{}
\theoremstyle{mystyle}
\renewcommand\figurename{Fig.~}

\newcommand{\nop}[1]{}

\makeatletter
\renewcommand{\@biblabel}[1]{[#1]\hfill}
\makeatother

\begin{document}

\hyphenpenalty=50000

\makeatletter
\newcommand\mysmall{\@setfontsize\mysmall{7}{9.5}}

\newenvironment{tablehere}
  {\def\@captype{table}}
  {}
\newenvironment{figurehere}
  {\def\@captype{figure}}
  {}

\thispagestyle{plain}%
\thispagestyle{empty}%

\begin{strip}
{\center
{\zihao{3}\textbf{
Principal Component Regression to Study the Impact of Economic Factors on Disadvantaged Communities}}
\vskip 9mm}

{\center {\sf \zihao{5}
Narmadha M. Mohankumar$^*$, Milan Jain, Heng Wan, Sumitrra Ganguli, \\
Kyle D. Wilson, and David M. Anderson\\

}

\vskip 3mm}
%

\centering{
\begin{tabular}{p{160mm}}

{\zihao{-5}
\linespread{1.6667} %
\noindent
\bf{Abstract:} {\sf
The Council on Environmental Quality’s Climate and Economic Justice Screening Tool defines "disadvantaged communities" (DAC) in the USA, highlighting census tracts where benefits of climate and energy investments are not accruing. We use a principal component generalized linear model, which addresses the intertwined nature of economic factors, income and employment and model their relationship to DAC status. Our study 
\begin{enumerate*}[label=(\arabic*)]
   \item identifies the most significant income groups and employment industries that impact DAC status
   \item provides the probability of DAC status across census tracts and compares the predictive accuracy with widely used machine learning approaches,
   \item obtains historical predictions of the probability of DAC status,
   \item obtains spatial downscaling of DAC status across block groups.
\end{enumerate*}
Our study provides valuable insights for policymakers and stakeholders to develop strategies that promote sustainable development and address inequities in climate and energy investments in the USA.}
}

\end{tabular}
}

\zihao{6}\end{strip}

\thispagestyle{plain}%
\thispagestyle{empty}%
\makeatother
\pagestyle{tstheadings}

\begin{figure}[b]
\vskip -6mm
\begin{tabular}{p{44mm}}
\\
\toprule\\
\end{tabular}
\\

\noindent
\setlength{\tabcolsep}{1pt}
\begin{tabular}{p{1.5mm}p{79.5mm}}
$\bullet$& Narmadha M. Mohankumar, Milan Jain, Heng Wan, Sumitrra Ganguli, Kyle D. Wilson, and David M. Anderson are with the  Pacific Northwest National Laboratory, Richland, Washington, USA. E-mail: \{narmadha.mohankumar, milan.jain, heng.wan, sumitrra.ganguli, kyle.wilson, dma\}@pnnl.gov\}\\
$\sf{*}$&
To whom correspondence should be addressed. 

\end{tabular}
\end{figure}\zihao{5}

\section{Introduction}
\noindent In response to Executive Order 14008, the United States government launched the Justice40 (J40) initiative in 2021 to address the impacts of climate change and environmental injustice which is currently incorporated into the Climate and Economic Justice Screening Tool (CEJST) \citep{CEJS}. The initiative aims to deliver 40 percent of the overall benefits of federal investments in climate and clean energy to disadvantaged communities (DACs), which are disproportionately impacted by environmental hazards, pollution, and other factors. To achieve this goal, the J40 initiative proposed criteria to identify which communities currently are not benefiting from climate and energy investments.

\begin{figure}[!b]
    \centering
    \includegraphics[width=\columnwidth, trim={0 0cm 0 0cm}, clip]{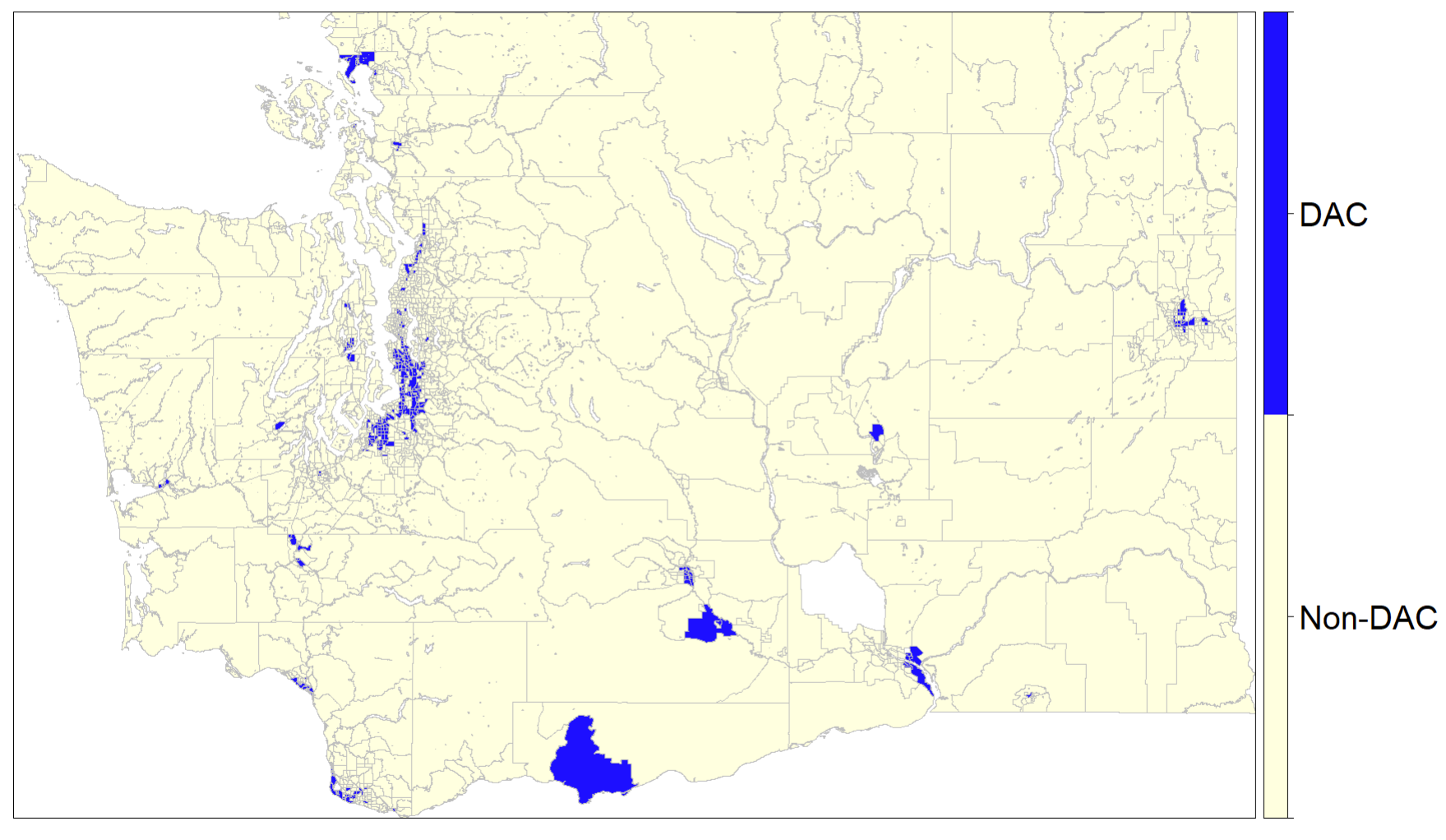}
    \caption{Distribution of DAC Communities in WA, USA in 2019.}
    \label{fig:wa}
\end{figure}

The J40 initiative uses a comprehensive set of criteria to determine DAC status involving socioeconomic vulnerabilities, environmental and climate hazards, energy burden, and fossil fuel dependence within a given geographic community \citep{dac_data}. Census tracts across the United States are the intended geographic resolution for this working definition and the DAC status designation from J40 is only available for the years after 2019. In the 2022c version of this working definition which is defined based on 2019 data, the DOE J40 initiative identified 15,172 census tracts across the United States as DACs, with 262 census tracts located in the state of Washington (WA), which is the primary focus of our study (Figure \ref{fig:wa}).

Recent studies have utilized DAC status to evaluate the effectiveness of policies and strategies in addressing the critical shortcomings of DAC communities and identifying opportunities for future investments. These studies have examined potential improvements in the digital-sharing economy, enterprise zone programs, and community development financial institutions to promote employment, income, reciprocity, social interaction, and resource accessibility \citep{vidal1995reintegrating,dillahunt2015promise, qian2020bikesharing}. It has been suggested that growing inequalities among communities may be attributed to factors such as the unavailability of low-cost housing, poor economic conditions, concentrations of minorities and female-headed families, and insufficient mental health care \citep{elliott1991structural,henderson2016influence}. Other studies have argued that disparities in the spatial accessibility of infrastructure are strongly associated with inequalities among communities and that equitable distribution of public and private sector investments in infrastructure projects is crucial \citep{leyshon1994access,brown2014spatial,mandarano2017equitable,wiesel2018locational}.  Furthermore, Ref. \cite{jain2023training} utilized applied machine learning (ML) models to predict the DAC status across census tracts for historical years. They used census data on income, employment, education, race, ethnicity, gender, and demographic information to predict DAC status. The findings from their study revealed that race and ethnicity played a significant role in predicting DAC status. 

Among the multitude of factors that influence DAC designation, the significance of economic factors such as income and employment cannot be overstated. These factors have a direct and profound impact on various aspects of DACs, including economic well-being, poverty alleviation, social mobility, community development, and the reduction of inequalities  \citep{osberg1985measurement,murali2004poverty,dillahunt2015promise,bilan2020impact, antipova2021unemployment}. While assessing the direct impact of these aspects on DAC status requires a substantial amount of data, such as information on health and healthcare access, inadequate infrastructure, transportation, technology access, crime rates, and social safety nets, it is important to acknowledge that comprehensive data across these areas may be lacking. In the absence of such comprehensive data, widely collected income and employment data can serve as valuable proxies to gain insights into the broader socio-economic challenges faced by disadvantaged communities \citep{osberg1985measurement,murali2004poverty,bilan2020impact}. 

Income and employment data is collected and analyzed across finer geographic units (e.g.,  block groups, blocks) and over many years \citep{ACS,lodes_data}, enabling a more precise assessment of disadvantaged communities and their specific economic challenges. By understanding the DAC distribution at finer geographic units, policymakers are better equipped to tailor strategies that address the unique economic circumstances of each local community, maximizing the impact of policies and programs aimed at improving economic conditions. By tracking DAC distribution changes over time, policymakers can assess the effectiveness of their interventions, identify areas where further action is required, and refine their strategies accordingly. Furthermore, by leveraging the projected income and employment data and forecasting the DAC distribution for future years, policymakers can make informed decisions and implement targeted interventions that effectively address the evolving economic conditions of disadvantaged communities.

Investigating the specific impacts of income and employment factors on DAC status is a challenging task due to their highly intercorrelated nature, influenced by numerous interrelated hidden factors that contribute to complex economic dynamics \citep{ray1998development, borjas2010labor}. In general, individuals who are employed tend to have higher incomes than those who are unemployed, and individuals with higher levels of education and experience tend to have higher-paying jobs. However, there are exceptions to this rule, such as individuals who work low-paying jobs despite having advanced degrees, or individuals who are unemployed but have significant savings or other sources of income. The correlation between income and employment can fluctuate based on various direct or indirect factors such as industry, geography, and economic conditions. Understanding these interrelated factors and identifying their individual impact on the DAC status requires a comprehensive approach that targets multiple dimensions simultaneously. 

Machine learning (ML) approaches have shown great potential in capturing complex relationships in data and providing high predictive accuracy. However, ML algorithms often operate as black-box models, meaning they do not explicitly reveal the functional relationships between input variables and model outcomes \citep{goebel2018explainable,lipton2018mythos}. This lack of interpretability can hinder a deeper understanding of the underlying dynamics. Furthermore, ML approaches typically do not incorporate uncertainty quantification in their estimates or predictions and may overlook the inherent variability in the data, which leads to challenges in assessing the variability or potential errors associated with the model’s outputs. In contrast, traditional models such as linear regression, logistic regression, and generalized linear models have been widely employed for decades in various fields of research to obtain interpretable results that allow for a deeper understanding of the relationship between the input variables and the outcome variable. These approaches provide estimated coefficients that indicate the magnitude and direction of the influence of each input variable on the outcome variable. Furthermore, these models provide uncertainty quantification for both the estimated coefficients for the relationship as well as for the predictions. However, these models are limited by assumptions about the linearity of the relationship and independence of the input variables and may struggle to accurately capture the underlying patterns and yield satisfactory performance in the presence of highly non-linear and complex relationships.

Principal component analysis (PCA) is a potential strategy for constructing effective models when dealing with highly non-linear and complex relationships while accounting for multicollinearity in the data. By employing PCA, the input variables are transformed into a potentially smaller set of uncorrelated components that capture the majority of the data's variation. This transformation simplifies the model by reducing the number of input variables and addresses issues related to multicollinearity \citep{jolliffe2016principal}. Principal component regression (PCR) is a powerful approach that combines PCA with regression models to model the relationship between input variables and the outcome variable \citep{kawano2018sparse,aguilera2006using}. In PCR, the PCA-derived components are used as input variables for the regression model. By accounting for multicollinearity in the data, PCR can provide reliable estimates for the relationship between the input variables and the outcome variable. 

\begin{figure}[!b]
    \centering
    \includegraphics[width=\columnwidth, trim={0 0cm 0 0cm}, clip]{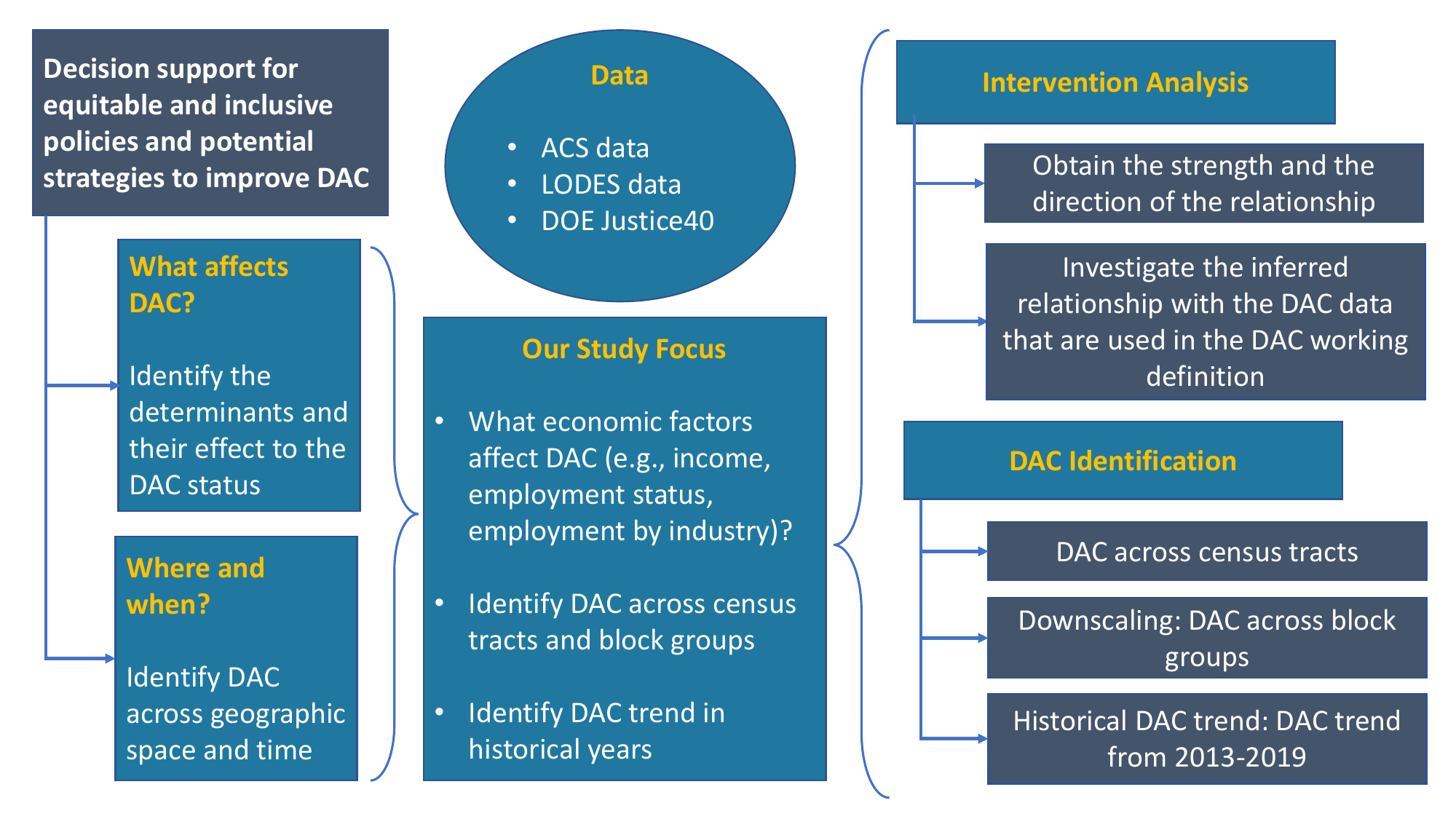}
    \caption{Overview of the study.}
    \label{fig:overview}
\end{figure}

Our study employs PCA and generalized linear models (GLMs) in the form of principal component generalized linear models (PCGLM) to comprehensively understand how economic factors such as income, employment status, and employment by industry affect the DAC status in Washington, USA. By employing PCGLM, we aim to model the effect of income, employment status, and employment by industry on the DAC status more comprehensively and predict the probability of DAC across different census geographic units (e.g., census tracts and block groups) and years. We gather data from three primary sources:  the American Community Survey (ACS) \citep{ACS}, the LEHD Origin-Destination Employment Statistics (LODES) \citep{lodes_data}, and the DOE Justice40 Initiative \citep{dac_data}.  By integrating data from these sources and applying our PCGLM, we 
\begin{enumerate*}[label=(\arabic*)]
    \item identify the most significant income groups and industries that impact the DAC status, 
    \item predict the likelihood of DAC across census tracts and compare our models' predictive performance with other machine learning approaches,  
    \item derive historical predictions of the probability of DAC status spanning from 2013 to 2018, and
    \item facilitate the spatial downscaling of DAC status by obtaining the probability of DAC status across block groups, enabling us to extend the analysis from census tracts to block groups (Figure \ref{fig:overview}). 
\end{enumerate*}
Our work provides an explicit understanding of the effect of income and employment on the DAC status, providing valuable insights regarding the economic distress within a community. DAC data from the J40 initiative lacks information on the DAC status before 2019 and identifying DAC status across census tracts in historical years is invaluable in understanding DAC transformations over time. Downscaling the DAC status to block groups offers a finer level of detail, capturing the spatial patterns and variability of DACs within a census tract. This finer level of detail provides localized information that is especially valuable for decision-making processes requiring a thorough understanding of unique challenges faced by communities in specific areas or regions within a census tract. These findings aid policymakers and stakeholders in making well-informed decisions to promote sustainable development and reduce economic disparities in disadvantaged communities and address inequities in climate and energy investments in the USA.

\section{Methodology}

\subsection{Data Collection}
\noindent We gathered data from three primary sources for our analysis. 

\begin{itemize}
\item  ACS: The American Community Survey (ACS) is a demographic survey program conducted by the United States Census Bureau which includes information about social, economic, demographic, and housing characteristics of the United States population across block groups \citep{ACS}.  For this study, we collected block-group level 16-binned household income from the ACS 5-year estimates 2015-2019. The 16-binned household income represents the population by 16 different income categories.
\item LODES: The Longitudinal Employer-Household Dynamics (LEHD) Origin-Destination Employment Statistics (LODES) dataset is a product of the United States Census Bureau's Longitudinal Employer-Household Dynamics (LEHD) program \citep{lodes_data}. LODES provides detailed job-related information by census blocks, including information on the number of workers and their earnings, industry of employment, and demographic characteristics. It is organized into three groups: (1) OD – Origin-Destination data associated with the transition of the employed population between home and work census blocks, (2) RAC – Residence area characteristics data by home census block, and (3) WAC – Workplace area characteristics data by work census block. For this study, we collected the number of workers by 20 industries of employment across census blocks from version LODES8 for the year 2019.
\item DAC: DAC data comes from the U.S. Department of Energy's working definition of disadvantaged communities as pertaining to \textsc{EO 14008}, or the Justice40 Initiative \citep{dac_data}, which is currently a part of the CEJST \citep{CEJS}. The DAC data includes thirty-six (36) burden variables collected at the census tract level and an indicator identifying each census tract as disadvantaged or not disadvantaged. The 36 indicators are taken from various data sources including ACS, LEHD Survey, and Environmental Justice Screening Tool (EJScreen), among others. For this study, we used the 2022c version of the DAC data \citep{CEJS}. For detailed information, refer to Justice40 DAC data documentation \citep{dac_data}. 
\end{itemize}

For this study, we aggregated the income and employment data from block groups and census blocks into census tracts, which is the target spatial scale for our models.

\subsection{Principal Component Generalized Linear Regression (PCGLM)}

\noindent We utilized PCR by combining PCA with GLM to model the DAC  status as a function of income, employment status, and employment by industry.  First, we applied PCA for the 38 variables related to income, employment status, and employment by industry. Through linear transformation, these variables are converted into principal components (PCs). The resulting PCs, or transformed variables, exhibit orthogonality and lack correlation with each other. This enables capturing the most important information contained within the original variables while mitigating the impact of multicollinearity. The presence of multicollinearity can adversely affect the coefficient estimates from the model leading to misleading conclusions about the relationship between input and outcome variables. 

Second, we employed GLM to model the relationship between the derived PCs and the DAC status. GLM is a flexible regression framework that can handle a variety of response distributions and link functions. In our study, we used a logit link with a binomial response distribution as
\begin{equation}
    logit(\boldsymbol{\pi})= \alpha_0 +\alpha_1 \textbf{z} _{1} +  \alpha_2 \textbf{z} _{2} + ... + \alpha_k \textbf{z} _{k} + \boldsymbol{\varepsilon}
\end{equation}
where, $\pi$ is the vector that represents the probability of DAC status across census tracts, $\textbf z_{1}, \textbf z_{2}, ..., \textbf z_{k}$ are the $k$ number of PCs, $\alpha_0, \alpha_1, ..., \alpha_k$ are the associated regression coefficients, and $\boldsymbol{\epsilon}$ is the random error term.  The PCs are linear combinations of the variables on income, employment status, and employment by industry such that,

\begin{equation}
\begin{aligned}
   \textbf z_{1}= l_{11} {\textbf x_1} +  l_{12} {\textbf x_2} + ... + l_{1p} {\textbf x_p} 
    \\
    \textbf z_{2}= l_{21} {\textbf x_1} +  l_{22} {\textbf x_2} + ... + l_{2p} {\textbf x_p} 
    \\
    .\\
    .\\
    .\\
\textbf z_{k}= l_{k1} {\textbf x_1} +  l_{k2} {\textbf x_2} + ... + l_{kp} {\textbf x_p} \\
\end{aligned}
\end{equation}

\noindent where, $\textbf x_{1}, \textbf x_{2}, ..., \textbf x_{p}$ represent the income, employment status, and employment by industry variables and $l_{11}, l_{12}, ..., l_{kp}$ represent the associated PC loadings. By employing PCGLM, we effectively combined the strengths of GLM and PCA to effectively model and understand how income, employment status, and employment by industry influence the DAC status.

\subsection{Variable Selection to Determine the Significant PCs to DAC Status}

\noindent In general, PCR uses the high-variance PCs in the model which correspond to higher eigenvalues, and discards the remaining low-variance components corresponding to lower eigenvalues. However,  these low-variance components can be as important as those with large variance to the response, thus, PCR may result in removing important PCs or including unimportant PCs in the model  \citep{jolliffe1982note}. To select the PCs for inclusion, we used a backward elimination approach. In this approach, we began with a GLM that included all of the PCs from the PCA and then eliminated the least informative PCs. Backward elimination eliminates redundant PCs from the GLM without appreciably increasing the residual sum of squares. This sequentially removes the least significant PC from the model until a desired level of significance is achieved or until no further PCs can be removed. This is effective in improving regression coefficient estimation while keeping the predictive power on the response. 

\subsection{Relationship of Income and Employment to DAC status}

\noindent By utilizing the significant PCs from backward elimination and by combining eq. (1) and eq.  (2), the PCGLM can be written as,

\begin{equation}
\begin{aligned}
    logit(\boldsymbol{\pi}) = & \alpha_0 + \alpha_1 (l_{11} {\textbf x_1} +  l_{12} {\textbf x_2} + ... + l_{1p} {\textbf x_p} ) +  \\&
    \alpha_2 (l_{21} {\textbf x_1} +  l_{22} {\textbf x_2} + ... + l_{2p} {\textbf x_p} ) + ... + \\&
    \alpha_k (l_{k1} {\textbf x_1} +  l_{k2} {\textbf x_2} + ... + l_{kp} {\textbf x_p} )+ \boldsymbol{\varepsilon} \\ = & \alpha_0 + (\alpha_1 l_{11} + \alpha_2 l_{21} + ...+ \alpha_k l_{k1}) {\textbf x_1} + \\&
    (\alpha_2 l_{12} + \alpha_2 l_{22} + ...+ \alpha_k l_{k2}){\textbf x_2} + ... + \\ & (\alpha_k l_{1p} + \alpha_2 l_{2p} + ...+ \alpha_k l_{kp}){\textbf x_p} + \boldsymbol{\varepsilon}\\ = &
    \alpha_0 + \beta_1 {\textbf x_1} + \beta_2{\textbf x_2} + ... +  \beta_p{\textbf x_p} + \boldsymbol{\varepsilon},
   \end{aligned}
\end{equation}

\noindent where $\beta_1, \beta_2, ..., \beta_p$ are the coefficients for the relationship of income, employment status, and employment by industry to the probability of DAC status. We reported the exponential value of the estimated coefficients ($e^{\hat{\beta}}$) for easy interpretation of the relationship. The coefficient estimates $e^{\hat{\beta}}$ represents the expected increase in the ratio of the probability of DAC to the probability of non-DAC status with a one-unit increase of the income and employment variables. We calculated the standard error ($\mathrm{SE}$) for the derived $e^{\hat{\beta}}$ by employing the delta method utilizing the Taylor series approximation \citep{papanicolaou2009taylor}. The 95\% confidence intervals for the coefficients are calculated by $e^{\hat{\beta}} \pm 1.96 \mathrm{SE}$. If 95\% CIs for $e^{\beta}$ do not include one, the associated variables are reported as significantly affecting the ratio of the probability of DAC to the probability of non-DAC status. Furthermore, we delved deeper into understanding the relationship between income, employment status, and employment by industry variables and the DAC status by closely examining their connection to the Justice40 DAC data. The Justice40 DAC data includes national percentiles of DAC indicators, which serve as criteria for defining the DAC status across census tracts. 

\subsection{Models' Predictive Accuracy}

\noindent We compared the predictive performance of our model with three commonly used machine learning approaches: Random Forest, Gradient Boosting, and Neural Nets. Although these approaches are typically not employed for inferring the relationship between a response and features, they are well-known for their ability to achieve high predictive accuracy by capturing complex patterns within the data. We evaluated the predictive accuracy of our proposed PCGLM  in predicting the DAC status across census tracts. To evaluate the models, we employed five-fold cross-validation during the model fitting process and calculated the F1-Score, precision, and recall for the test data. The F1-Score represents the harmonic mean of the precision and recall values that are obtained from applying the models to the test data. We reported the F1-Score, precision, and recall and compared the predictive accuracy of the models in predicting the DAC status and non-DAC status. 

In contrast to ML approaches, PCGLM offers the advantage of providing uncertainty quantification for predictions related to the probability of DAC status. This is achieved by calculating the 95\% prediction interval for the predicted probability for each census tract. By calculating the 95\% prediction interval, we identified the prediction uncertainty and determined which census tracts exhibit higher uncertainty in the predicted probability of DAC status. Additionally, we compared the prediction uncertainty between DAC and non-DAC census tracts by calculating the average 95\% prediction interval separately for each DAC and non-DAC category. 

\subsection{Historical Predictions of DAC Status}

\noindent The coefficients derived from our PCGLM allow us to infer the relationship between income, employment status, and employment by industry to DAC status using data collected in 2019. Identifying the DAC status across census tracts in previous years is invaluable in understanding DAC transformations over time. To achieve this, we utilized data on income, employment status, and employment by industry sourced from the ACS and LODES data spanning from 2013 to 2018. By employing this data in our PCGLM, we obtained the predicted probabilities of DAC status across census tracts in WA from 2013 to 2018, providing a comprehensive analysis of historical DAC trends across prior years to 2019.

\subsection{Downscaling: Prediction of DAC Status across Block Groups}

\noindent We utilized our proposed PCGLM to generate downscaled predictions of DAC status at the block group level. In our model, we incorporated data on income, employment status, and employment by industry specific to each block group in WA and predicted the probability of DAC across block groups within the state. To evaluate the accuracy of our predictions, we compared the predicted DAC status of block groups with the true DAC status of their corresponding census tracts. In this comparison, we performed two evaluations. Firstly, we calculated the mean probability of DAC status for block groups within each census tract. Based on this mean probability, we classified the census tract as either DAC or non-DAC. We then calculated the F1-score to assess the accuracy of the DAC status predictions for the census tracts.

\section{Results}

\subsection{Relationship of Income and Employment to DAC status}

\begin{figure}[!b]
  \centering
  \includegraphics[width=\linewidth,clip]{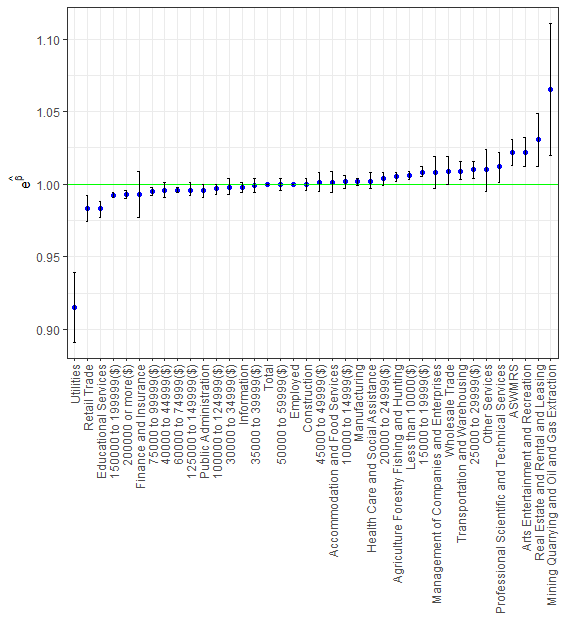}
  \caption{Exponential of the coefficient estimates ($\hat{\boldsymbol\beta}$) of income, employment, and employment by industry obtained from our PCGLM.}
   \label{fig:estimates}
\end{figure}

\noindent The strength of the estimated relationship ($e^{\hat{\beta}}$) and the 95\% CIs are displayed in Table \ref{table1} and Figure \ref{fig:estimates}. The 95\% CIs in Table \ref{table1} indicated that the mining quarrying and oil and gas extraction industry and the utilities industry have the highest impact on the ratio of the probability of DAC to the probability of non-DAC status (Table \ref{table1} and Figure \ref{fig:estimates}). Specifically, the mining quarrying and oil and gas extraction industry showed a positive effect, while the utilities industry showed a negative effect on the probability of DAC to the probability of non-DAC ratio. However, coefficients for the mining quarrying and oil and gas extraction industry and the utilities industry yielded wider 95\% CIs compared to other income and employment variables indicating a comparatively higher uncertainty in the estimated coefficients. Other industries that yielded a positive significant effect are agriculture, forestry, fishing and hunting, wholesale trade, transportation and warehousing, real estate and rental and leasing, professional scientific and technical services, administrative, support, waste management, and remediation services (ASWMRS), and arts entertainment and recreation.  Other industries apart from the utilities industry that yielded a negative significant effect are retail trade and educational services.  The inferred effect of income bins on DAC status showed that low-income bins have a positive significant effect on the ratio of the probability of DAC to the probability of non-DAC status, and high-income bins have a negative significant effect. The strength of these estimated relationships is displayed in Figure \ref{fig:estimates}. 

\begin{figure}[!b]
  \centering
  \includegraphics[width=\linewidth, trim={0cm 0 0 0cm}, clip]{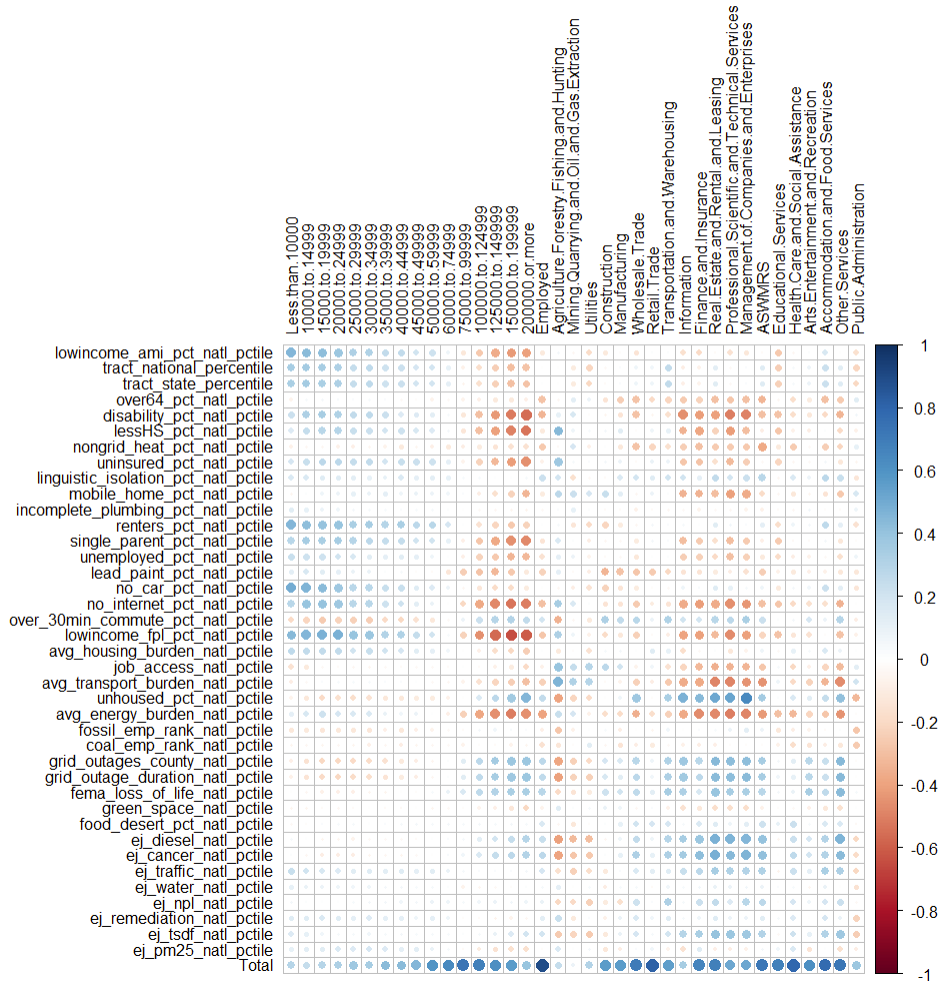}
  \caption{Correlation plot showing Pearson's correlation among the variables of income, employment, and the Justice40 DAC data. DAC working definition is derived using the Justice40 DAC data. }
   \label{fig:corr}
\end{figure}

Pearson's correlation plot in Figure \ref{fig:corr} shows the correlation between income and employment features to Justice40 DAC data involving national percentiles that are used in the working definition of DAC status. Among the most significant four industries that showed a positive impact on DAC status, mining quarrying and oil, and gas extraction showed a positive correlation with transportation burden. The real estate and rental and leasing employments exhibited a positive correlation with multiple DAC variables, including unhoused individuals, grid outages, and environmental justice concerns related to diesel pollution and cancer rates. The arts, entertainment, and recreation industry showed a positive correlation with the FEMA loss of life percentile. The administrative, support, waste management, and remediation services industry (ASWMRS) exhibited a positive correlation with environmental justice concerns related to diesel pollution and cancer rates. 

Among the most significant three industries that showed a negative impact on DAC status, the utilities industry demonstrated a negative correlation with environmental justice concerns regarding diesel pollution and cancer rates. Employment in educational services showed negative correlations with energy burden, low income, no internet, single parent, less HS, and disability. Employment in the retail trade industry demonstrated a negative correlation with lead paint concerns. Overall, the correlation provides valuable insights into the hidden factors that may be associated with employment and industry that may contribute to the DAC status. Understanding these relationships can aid in identifying how significant impacts of income and employment on DAC status are related to other areas of focus for policy and intervention. However, it is important to note that these correlations don’t imply causation and deeper analysis is required to identify true causes. 

\begin{table}
  \caption{Exponential of the coefficient estimates ($\hat{\boldsymbol\beta}$) with 95\% confidence intervals (CIs). Gray shaded rows indicate that the variable has a positive significant effect on the ratio of  probability of DAC to probability of non-DAC, and the blue shaded rows indicate a negative significant effect.}
\centering
\vspace{0.5cm}
\setlength\extrarowheight{1.5pt}
\begin{tabular}{ p{5.8cm} p{0.5cm} p{1.5cm}  }
\hline
Features of income, employment and jobs & \textbf{$e^{\hat{\boldsymbol\beta}}$} & 95\% CI \\ 
\midrule
Total population                        &                  1.000  & (1.000,1.000) \\
\rowcolor{lightgray} Income (\$): Less than 10000           &                         1.006  & (1.003,1.009)\\
Income (\$): 10000 to 14999          &                          1.002  &    (0.997,1.006)\\
\rowcolor{lightgray} Income (\$): 15000 to 19999        &                            1.008 &     (1.005,1.012)\\
Income (\$): 20000 to 24999        &                            1.004 &    (0.999,1.008)\\
\rowcolor{lightgray} Income (\$): 25000 to 29999        &                            1.010 &    (1.004,1.016)\\
Income (\$): 30000 to 34999               &                     0.998 &    (0.993,1.004)\\
Income (\$): 35000 to 39999                      &              0.999 &    (0.994,1.004)\\
Income (\$): 40000 to 44999                     &               0.996 &    (0.991,1.001)\\
Income (\$): 45000 to 49999                        &            1.001 &    (0.995,1.008)\\
Income (\$): 50000 to 59999                          &          1.000 &     (0.996,1.004)\\
\rowcolor{LightCyan} Income (\$): 60000 to 74999  &                                   0.996 &    (0.994,0.998)\\
\rowcolor{LightCyan} Income (\$): 75000 to 99999      &                              0.995 &    (0.992,0.998)\\
\rowcolor{LightCyan} Income (\$): 100000 to 124999        &                          0.997 &    (0.993,1.000)\\
Income (\$): 125000 to 149999          &                        0.996 &    (0.992,1.001)\\
\rowcolor{LightCyan} Income (\$): 150000 to 199999             &                     0.992 &    (0.991,0.994)\\
\rowcolor{LightCyan} Income (\$): 200000 or more                    &                0.993 &    (0.990,0.996)\\
Employed population                                 &          1.000 &    (1.000,1.000)\\
\rowcolor{lightgray} Agriculture, forestry, fishing, and hunting   &        1.005 &    (1.002,1.008)\\
\rowcolor{lightgray} Mining, quarrying, and oil and gas extraction  &      1.065 &    (1.020,1.111)\\
\rowcolor{LightCyan}Utilities                                  &        0.915 &     (0.891,0.939)\\
Construction                          &             1.000 &    (0.996,1.004)\\
Manufacturing                           &           1.002 &    (0.999,1.004)\\
\rowcolor{lightgray} Wholesale trade                            &        1.009 &    (1.000,1.019)\\
\rowcolor{LightCyan} Retail trade                              &         0.983 &     (0.974,0.992)\\
\rowcolor{lightgray} Transportation and warehousing     &                1.009 &    (1.003,1.016)\\
Information                               &         0.998 &    (0.994,1.001)\\
Finance and insurance                  &            0.993 &    (0.977,1.009)\\
\rowcolor{lightgray} Real estate and rental and leasing    &             1.031 &    (1.012,1.049)\\
\rowcolor{lightgray} Professional scientific and technical services  &    1.012 &    (1.001,1.022) \\
Management of companies and enterprises      &      1.008 &    (0.997,1.019)\\
\rowcolor{lightgray} ASWMRS                                       &      1.022 &    (1.013,1.031)\\
\rowcolor{LightCyan} Educational services                        &       0.983 &    (0.977,0.988)\\
Health care and social assistance       &           1.002 &    (0.997,1.008)\\
\rowcolor{lightgray} Arts entertainment and  recreation           &       1.022 &    (1.012,1.032)\\
Accommodation and food services              &      1.001 &    (0.994,1.009)\\
Other services                             &        1.010 &    (0.995,1.024)\\
Public administration                 &             0.996 &    (0.991,1.000)\\
Longitude                                  &               0.794 &    (0.31,1.278)\\
Area                               &              1.000 &    (0.652,1.348)\\
Intercept                           &               0.000 &    (0.000,0.000)\\

\bottomrule
\end{tabular}
 \label{table1}
\end{table}

\subsection{Predictions of DAC Status across Census Tracts}

\noindent In Table \ref{table2}, we present the F1 score, precision, and recall metrics for the test data set, allowing for a comparison between our PCGLM and other machine learning approaches. The higher F1 score for the neural net indicated that the neural net achieved the best predictive accuracy in predicting both DAC and Non-DAC. All approaches yielded comparable F1 scores, however, our PCGLM provided similar predictive accuracy to neural net in predicting non-DAC, and second best in predicting DAC. Nevertheless, the F1 score for DAC and non-DAC across all approaches indicates that their performance in predicting DAC status is relatively lower compared to predicting non-DAC status.

\begin{figure}
  \centering
  \includegraphics[width=\linewidth,trim={0 0 0 0},clip]{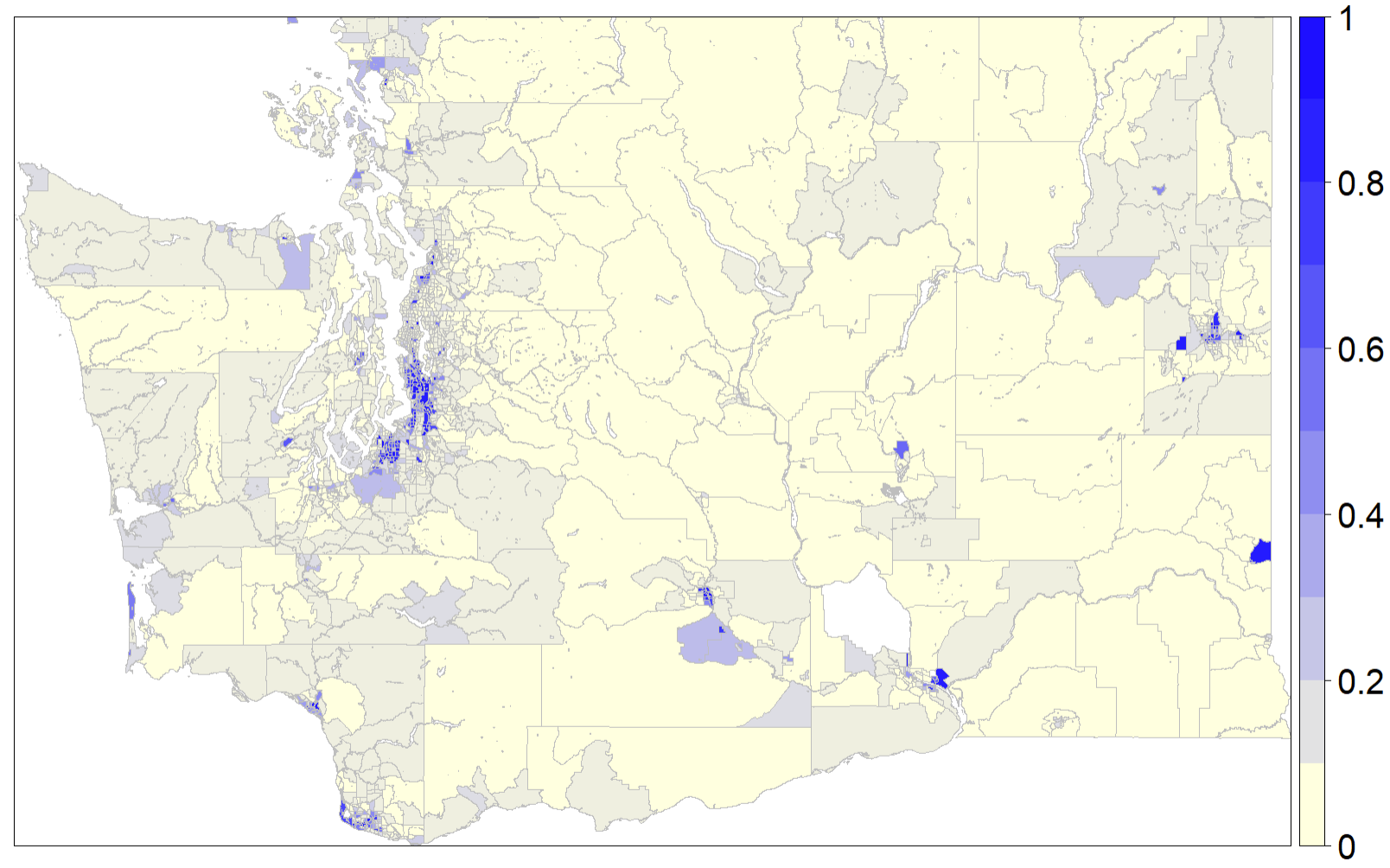}
  \caption{Predicted probability of DAC status across census tracts in WA, USA in 2019.}
   \label{fig:predcensus}
\end{figure}

Figure \ref{fig:predcensus} shows the predicted probability of DAC status generated by our PCGLM for the census tracts in Washington State in 2019. The average 95\% prediction interval for the probability of DAC within the census tracts that are defined as DAC is 0.12, while it is 0.30 for census tracts that are defined as non-DAC. Consistent with the findings from the F1 scores, this suggests that our model's performance in predicting DAC status is relatively lower compared to predicting non-DAC status.

\begin{figure*}
  \centering
  \includegraphics[width=\linewidth, trim={0 5cm 0 3cm},clip]{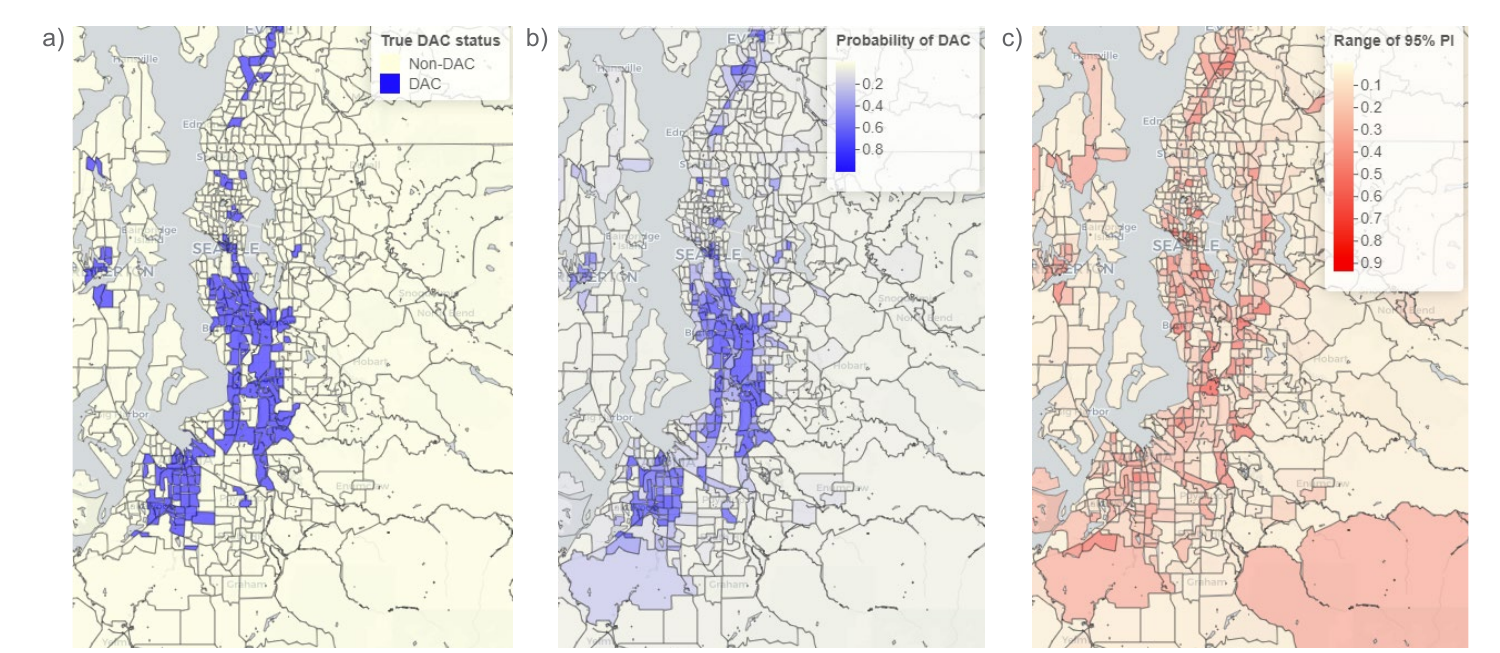}
\caption{ a) True DAC status across census tracts in Seattle region in 2019, b) Predicted probability of DAC status across census tracts in Seattle region in 2019, c) 95\% prediction interval for the probability of DAC status across census tracts in Seattle region.}
   \label{fig:predseat}
\end{figure*}

Figure \ref{fig:predseat} displays the predicted probability of DAC status across census tracts in the Seattle region, providing a comparison with the true DAC status in those census tracts. Additionally, we show the 95\% prediction interval for the probability of DAC status across census tracts, highlighting the census tracts with the highest predictive uncertainty.

\begin{table*}[!b]
\centering
  \caption{Comparison on models' predictive accuracy}
  \label{tab:freq}
\vspace{0.5cm}
  \begin{tabular}{p{0.1\textwidth}>
{\centering}p{0.13\textwidth}>
{\centering}p{0.13\textwidth}>
{\centering}p{0.13\textwidth}>
{\centering}p{0.13\textwidth}>
{\centering}p{0.13\textwidth}>
{\centering\arraybackslash}p{0.14\textwidth}}
\hline
    Model & F1-score (Non-DAC) & F1-score (DAC) & Precision (Non-DAC) & Precision (DAC) & Recall (Non-DAC) & Recall (DAC) \\
\hline
    PCGLM & 0.945 & 0.805 & 0.918 & 0.842 & 0.975 & 0.608\\
    RF & 0.943 & 0.672 & 0.906 & 0.878 & 0.983 & 0.544\\
    GBM & 0.944 & 0.692 & 0.913 & 0.852 & 0.978 & 0.582\\
    NNet & 0.945 & 0.817 & 0.941 & 0.763 & 0.949 & 0.734\\
  \hline
\end{tabular}
 \label{table2}
\end{table*}

\subsection{Historical Predictions of the Probability of DAC Status}

\begin{figure*}[!b]
  \centering
  \includegraphics[width=\linewidth,trim={0 0cm 0 0cm},clip]{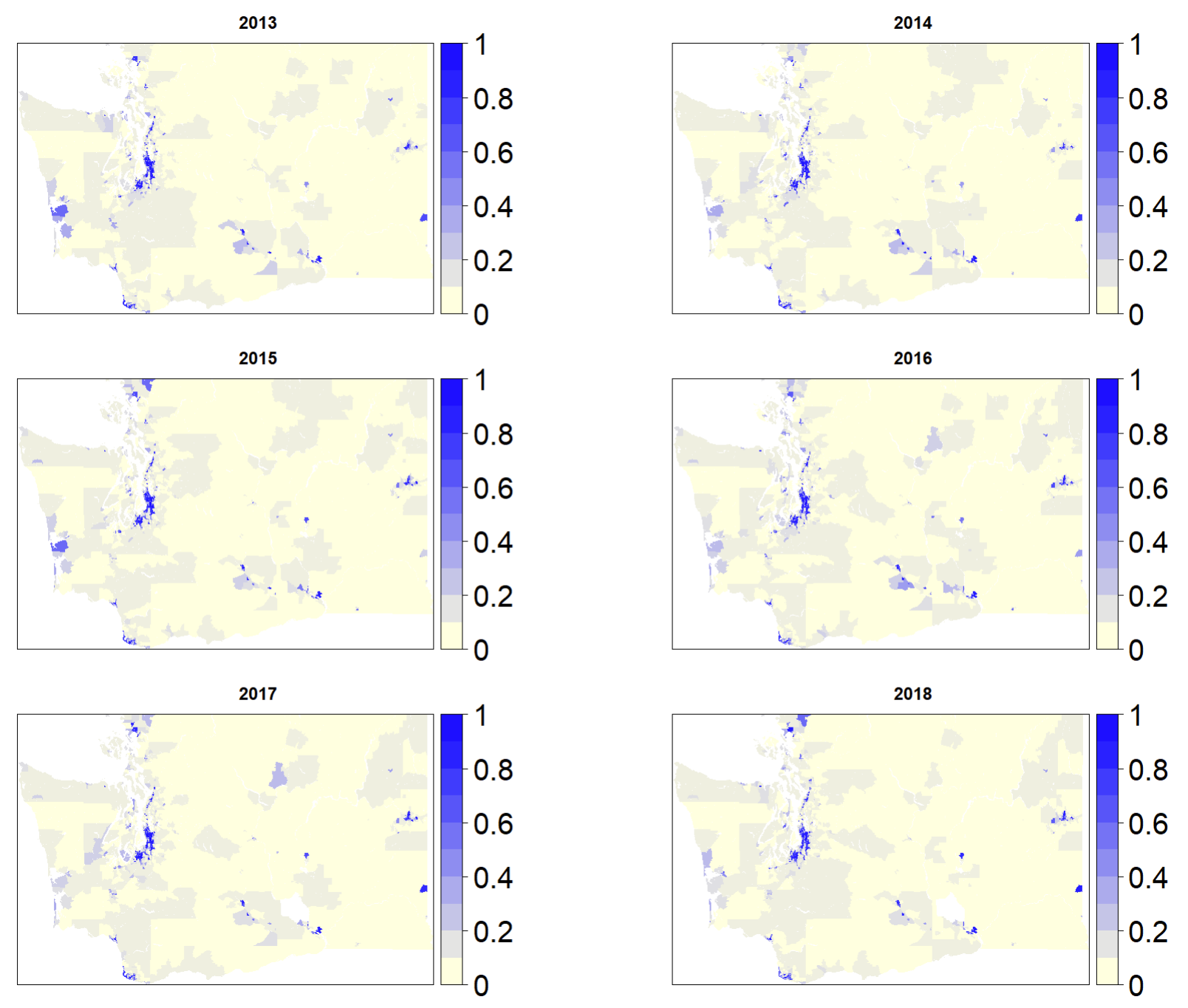}
  \caption{Predicted probability of DAC status across census tracts in WA, USA from 2013 to 2018.}
   \label{fig:predhist}
\end{figure*}

\noindent Historical predictions of the probably of DAC status from 2013 to 2018 are shown in  Figure \ref{fig:predhist}. Our findings indicate a progressive decline in the DAC communities from 2013 to 2018, with the percentage of DAC status predicted at 17.45\%, 17.09\%, 17.71\%, 16.61\%, 16.27\%, and 16.89\% for the respective years. We also noticed that throughout the years the disadvantaged census tracts have a similar distribution. 

\subsection{Downscaling: Prediction of DAC status Across Block Groups}

\begin{figure}
  \centering
  \includegraphics[width=0.98\linewidth,clip]{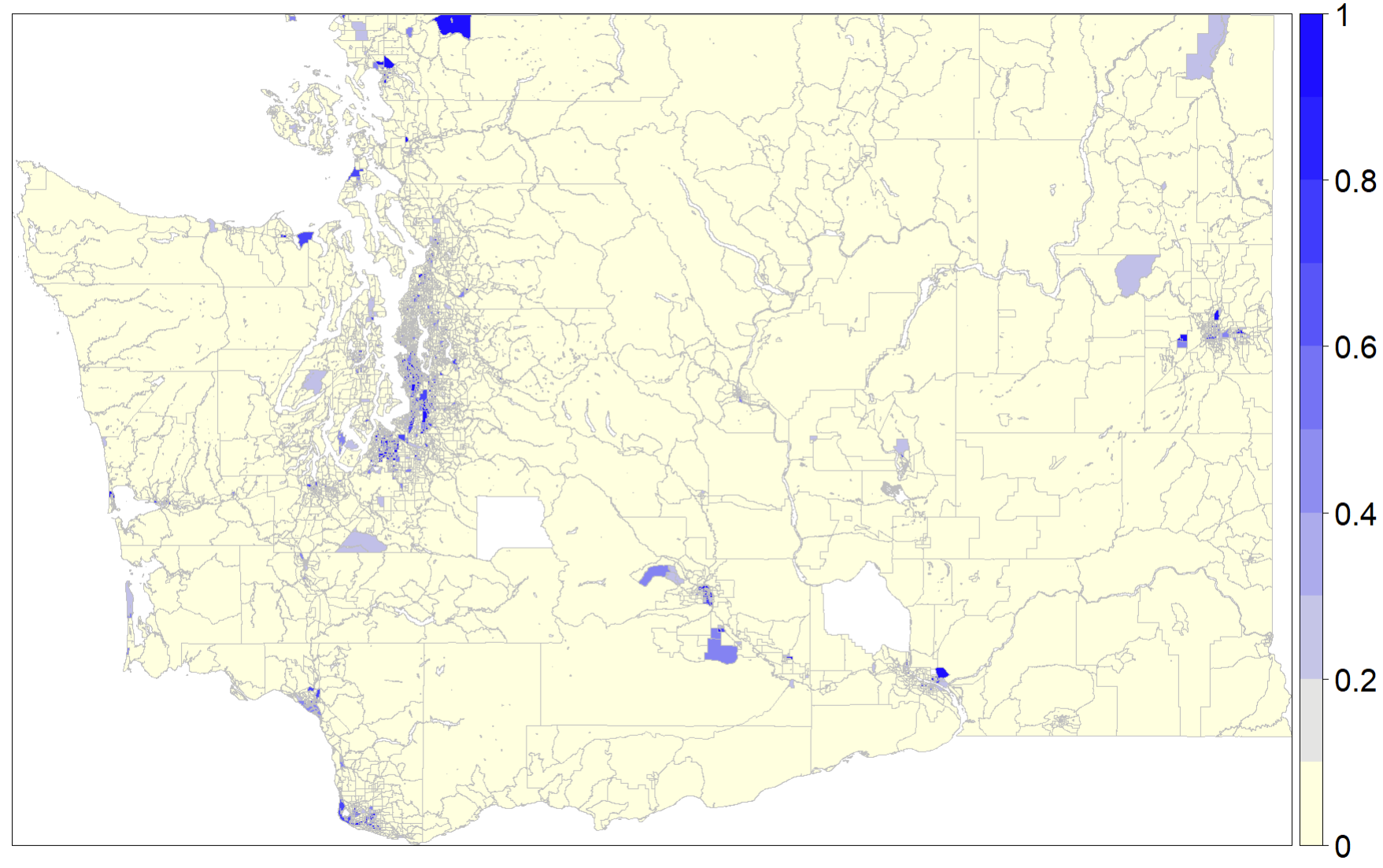}
  \caption{Predicted probability of DAC status across block groups in WA, USA in 2019.}
   \label{fig:predblk}
\end{figure}

\begin{figure*}
  \centering
  \includegraphics[width=0.97\linewidth,trim={0 4cm 0 0},clip]{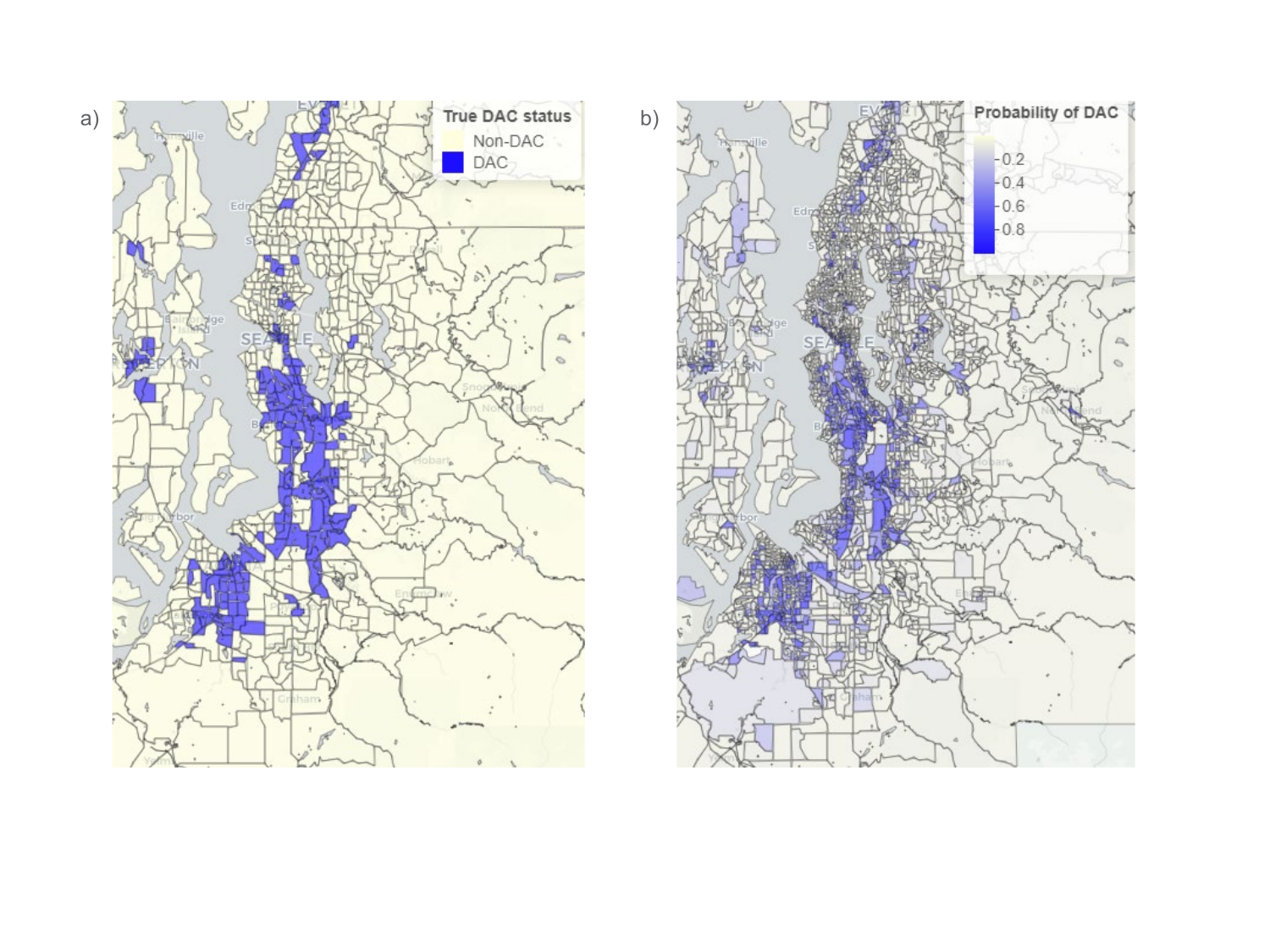}
\caption{ a) True DAC status across census tracts in Seattle region in 2019, b) Predicted probability of DAC status across block groups in Seattle region in 2019. }
   \label{fig:predseatblk}
\end{figure*}

\noindent Figure \ref{fig:predblk} represents the predicted probability of DAC status across block groups in WA which is obtained using block-group level ACS and LODES data in our PCGLM.  In our model evaluations, the F1-score for the DAC status predictions of census tracts, obtained by averaging the block group level predictions was 0.94 for predicting DAC and 0.67 for predicting non-DAC. Additionally, in Figure \ref{fig:predseatblk}, we present the predicted probability of DAC status across block groups in the Seattle region, along with a comparison to the true DAC status across census tracts in the region. Overall, our block group level predictions on the probability of DAC closely align with the true DAC status of the census tracts.

\section{Discussion and Conclusion}

\noindent Our PCGLM approach serves two main objectives in studying the DAC status. First, it addresses multicollinearity in the data, allowing us to establish explicit functional relationships between factors related to income, employment status, employment by industry, and DAC status. The high multicollinearity among input variables could be attributed to the positive correlation observed in most of the income, employment status, and employment by industry, which is important to account for to accurately estimate the explicit functional relationships between these factors and the DAC status. Second, our approach builds a predictive model that can assess the probability of DAC status spatially and temporally.  Our comparison of the PCGLM with other widely used ML techniques shows how the predictive accuracy of PCGLM compares to the ML approaches that are widely known to provide higher predictive accuracy. Additionally, our PCGLM provides uncertainty quantification for both the estimated coefficients representing the relationship between income, employment status, employment by industry, and the DAC status, as well as for the predictions regarding the probability of the DAC status. This feature distinguishes PCGLM from many other machine learning methods, which typically do not incorporate uncertainty quantification in their estimates. The predicted probability of DAC status across historical years from 2013 to 2018 reveals the trends in DAC communities across census tracts over time, offering insights into the transformation of DAC communities. The predicted probability of DAC status across block groups allows for downscaling the DAC status to a finer spatial resolution beyond census tracts, providing a high-level understanding of the distribution of DAC communities at the block group level. 
Overall, our approach offers insights into the impact of economic factors on the DAC status and temporal trends and spatial distribution of DAC communities at both the census tract and block group levels.

Our insights from the functional relationship between employment by industry to the DAC status indicate that increasing employment in utilities, educational services, and retail and trade industries can be advantageous in enhancing the well-being of disadvantaged communities. Based on the relationship of the significant income and employment factors to the DAC variables, the hidden factors that may contribute to the DAC status include transportation burden, unhoused individuals,  grid outages, grid outage duration, and environmental justice concerns related to diesel pollution and cancer rates (see Figure \ref{fig:estimates} and Figure \ref{fig:corr}). These are potential areas to be considered in policy approaches for designing equitable and inclusive policies for disadvantaged communities. 

The predictive accuracy of our PCGLM and other widely used ML approaches demonstrates proficiency in predicting non-DAC status across census tracts. However, when it comes to predicting DAC status, their performance is comparatively lower. This suggests that the data available for DAC census tracts may not provide sufficient information or the variability in the data are higher to accurately predict DAC status using these models. One reason for this limitation is that the DAC working definition, derived from the Justice40 initiative, incorporates factors such as socioeconomic vulnerabilities, environmental and climate hazards, energy burden, and fossil fuel dependencies. Unfortunately, we do not have data available to include these factors in our model. Moreover, among all the census tracts in data, approximately 20\% of them are categorized as DAC. The income and employment data gathered from these DAC census tracts exhibit significant variability. As a result, our models may not fully capture the complexities and nuances that contribute to predicting DAC status. Furthermore, the ACS data is subject to a notable amount of uncertainty and this inherent uncertainty can further impact the predictive accuracy of the models, particularly when trying to assess the DAC status at fine spatial scales. However, it is worth highlighting that our study leverages the available data on income and employment, which are collected at a higher spatial resolution across many years for various purposes. Income and employment have a direct and profound impact on various aspects of disadvantaged communities, including economic well-being, poverty alleviation, social mobility, community development, and the reduction of inequalities. Therefore, using income and employment as proxies for studying disadvantaged communities provides valuable insights into the broader socio-economic challenges faced by the DAC, even in cases where specific data used for evaluating the DAC status may be lacking. In summary, while the predictive accuracy of models for DAC status prediction may be limited due to the lack of comprehensive data on specific factors and the uncertainty in ACS data, our study takes advantage of income and employment data collected at higher spatial resolutions over the years enabling us to capture some of the contributing factors to the DAC and investigate the DAC distribution. Furthermore, by utilizing projected income and employment data for future years and forecasting the DAC distribution, policymakers can make informed decisions and implement targeted interventions that effectively address the evolving economic conditions of disadvantaged communities.

In Ref. \cite{jain2023training}'s work, they incorporated education, race, ethnicity, gender, and demographic information alongside income and employment to predict DAC status. The findings from this study revealed that race and ethnicity played a significant role in determining DAC status, often overshadowing the impact of income and employment. Notably, the model was projecting census tracts with a high African-American population as DAC. Furthermore, findings from this study indicated a potential decrease in the total number of DAC communities from 2013 to 2019 which is similar to the finding from our study. Our current research focuses on this prior study to develop an explainable model that gives explicit functional relationships between income and employment to DAC status. Our future work focuses on incorporating other potential factors and assessing their correlations as well as causal effects to improve our findings regarding the DAC and the determinants of DAC. 

In our PCGLM approach, we employ variable selection techniques to identify the significant PCs. Specifically, we utilize backward elimination, a step-wise procedure that iteratively removes the least significant PC from the model until reaching a desired level of significance. This method effectively enhances the estimation of regression coefficients while preserving the predictive power of the model. However, one can consider utilizing modern penalized regression techniques like LASSO (Least Absolute Shrinkage and Selection Operator) and ridge regression \citep{tibshirani1996regression}. These techniques are particularly well-suited for handling high-dimensional data, which often introduces the risk of overfitting. Overfitting occurs when a model performs exceptionally well on the training data but struggles to generalize to new data samples. Modern penalized regression techniques can effectively handle high-dimensional data by appropriately shrinking or constraining the coefficients associated with less relevant variables. This regularization helps mitigate the overfitting problem, leading to improved generalization performance on new data. 

Our model's block group level predictions offer valuable insights into the probability of DAC status with a higher spatial resolution. The findings from our study emphasize the significance of recognizing the variability within census tracts and the importance of conducting a more granular assessment when making policy decisions related to DAC communities at the block group level. Our study highlights that even if a census tract is officially classified as a non-DAC, there may still be block groups within that census tract that exhibit DAC characteristics. This indicates the presence of localized disparities and reinforces the need to consider the heterogeneity within census tracts when formulating policies and interventions targeting DAC communities. By providing a more detailed understanding of DAC probabilities at the block group level, our study enhances the precision and accuracy of decision-making processes, allowing policymakers to better identify and address specific areas and populations that require targeted support and resources.

\section*{Acknowledgment}
\noindent This research was supported by the Agile Initiative, a multi-disciplinary Pacific Northwest National Laboratory (PNNL) initiative. PNNL is operated by Battelle Memorial Institute under Contract DE-AC06-76RL01830.

\renewcommand\refname{\zihao{10}\textbf{References}}

\bibliographystyle{apa}
\bibliography{manuscript_with_author}

\begin{thebibliography}{28}
\expandafter\ifx\csname natexlab\endcsname\relax\def\natexlab#1{#1}\fi
\providecommand{\url}[1]{\texttt{#1}}
\providecommand{\href}[2]{#2}
\providecommand{\path}[1]{#1}
\providecommand{\DOIprefix}{doi:}
\providecommand{\ArXivprefix}{arXiv:}
\providecommand{\URLprefix}{URL: }
\providecommand{\Pubmedprefix}{pmid:}
\providecommand{\doi}[1]{\href{http://dx.doi.org/#1}{\path{#1}}}
\providecommand{\Pubmed}[1]{\href{pmid:#1}{\path{#1}}}
\providecommand{\bibinfo}[2]{#2}
\ifx\xfnm\relax \def\xfnm[#1]{\unskip,\space#1}\fi
\bibitem[{Aguilera et~al.(2006)Aguilera, Escabias \&
  Valderrama}]{aguilera2006using}
\bibinfo{author}{Aguilera, A.~M.}, \bibinfo{author}{Escabias, M.}, \&
  \bibinfo{author}{Valderrama, M.~J.} (\bibinfo{year}{2006}).
\newblock \bibinfo{title}{Using principal components for estimating logistic
  regression with high-dimensional multicollinear data}.
\newblock {\it \bibinfo{journal}{Computational Statistics \& Data Analysis}\/},
   {\it \bibinfo{volume}{50}\/}, \bibinfo{pages}{1905--1924}.
\bibitem[{Antipova \& Momeni(2021)}]{antipova2021unemployment}
\bibinfo{author}{Antipova, A.}, \& \bibinfo{author}{Momeni, E.}
  (\bibinfo{year}{2021}).
\newblock \bibinfo{title}{Unemployment in socially disadvantaged communities in
  {T}ennessee, {US}, during the covid-19}.
\newblock {\it \bibinfo{journal}{Frontiers in Sustainable Cities}\/},  {\it
  \bibinfo{volume}{3}\/}, \bibinfo{pages}{726489}.
\bibitem[{Bilan et~al.(2020)Bilan, Mishchuk, Samoliuk \&
  Yurchyk}]{bilan2020impact}
\bibinfo{author}{Bilan, Y.}, \bibinfo{author}{Mishchuk, H.},
  \bibinfo{author}{Samoliuk, N.}, \& \bibinfo{author}{Yurchyk, H.}
  (\bibinfo{year}{2020}).
\newblock \bibinfo{title}{Impact of income distribution on social and economic
  well-being of the state}.
\newblock {\it \bibinfo{journal}{Sustainability}\/},  {\it
  \bibinfo{volume}{12}\/}, \bibinfo{pages}{429}.
\bibitem[{Borjas \& Van~Ours(2010)}]{borjas2010labor}
\bibinfo{author}{Borjas, G.~J.}, \& \bibinfo{author}{Van~Ours, J.~C.}
  (\bibinfo{year}{2010}).
\newblock {\it \bibinfo{title}{Labor economics}\/}.
\newblock \bibinfo{publisher}{McGraw-Hill/Irwin Boston}.
\bibitem[{Brown \& Lloyd-Jones(2014)}]{brown2014spatial}
\bibinfo{author}{Brown, A.}, \& \bibinfo{author}{Lloyd-Jones, T.}
  (\bibinfo{year}{2014}).
\newblock \bibinfo{title}{Spatial planning, access and infrastructure}.
\newblock In {\it \bibinfo{booktitle}{Urban Livelihoods}\/} (pp.
  \bibinfo{pages}{211--227}).
\newblock \bibinfo{publisher}{Routledge}.
\bibitem[{{Census Bureau}(2022{\natexlab{a}})}]{ACS}
\bibinfo{author}{{Census Bureau}, U.} (\bibinfo{year}{2022}{\natexlab{a}}).
\newblock {\it \bibinfo{title}{{American Community Survey} {5-Years
  Estimates}}\/}.
\newblock \URLprefix
  \url{https://www.census.gov/data/developers/data-sets/acs-5year.html}.
\bibitem[{{Census Bureau}(2022{\natexlab{b}})}]{lodes_data}
\bibinfo{author}{{Census Bureau}, U.} (\bibinfo{year}{2022}{\natexlab{b}}).
\newblock {\it \bibinfo{title}{Longitudinal Employer-Household Dynamics (LEHD)
  Survey}\/}.
\newblock \URLprefix \url{https://lehd.ces.census.gov/data/}.
\bibitem[{{Council on Environmental Quality}(2022)}]{CEJS}
\bibinfo{author}{{Council on Environmental Quality}} (\bibinfo{year}{2022}).
\newblock \bibinfo{title}{Methodology \& data - climate \& economic justice
  screening tool}.
\newblock
  \bibinfo{note}{\url{https://screeningtool.geoplatform.gov/en/methodology}.
  Accessed: 2023-7-24}.
\bibitem[{Dillahunt \& Malone(2015)}]{dillahunt2015promise}
\bibinfo{author}{Dillahunt, T.~R.}, \& \bibinfo{author}{Malone, A.~R.}
  (\bibinfo{year}{2015}).
\newblock \bibinfo{title}{The promise of the sharing economy among
  disadvantaged communities}.
\newblock In {\it \bibinfo{booktitle}{proceedings of the 33rd annual ACM
  conference on human factors in computing systems}\/} (pp.
  \bibinfo{pages}{2285--2294}).
\bibitem[{DOE(2022)}]{dac_data}
\bibinfo{author}{DOE} (\bibinfo{year}{2022}).
\newblock {\it \bibinfo{title}{Justice40 DAC Data}\/}.
\newblock \URLprefix \url{https://energyjustice.egs.anl.gov/}.
\bibitem[{Elliott \& Krivo(1991)}]{elliott1991structural}
\bibinfo{author}{Elliott, M.}, \& \bibinfo{author}{Krivo, L.~J.}
  (\bibinfo{year}{1991}).
\newblock \bibinfo{title}{Structural determinants of homelessness in the united
  states}.
\newblock {\it \bibinfo{journal}{Social problems}\/},  {\it
  \bibinfo{volume}{38}\/}, \bibinfo{pages}{113--131}.
\bibitem[{Goebel et~al.(2018)Goebel, Chander, Holzinger, Lecue, Akata, Stumpf,
  Kieseberg \& Holzinger}]{goebel2018explainable}
\bibinfo{author}{Goebel, R.}, \bibinfo{author}{Chander, A.},
  \bibinfo{author}{Holzinger, K.}, \bibinfo{author}{Lecue, F.},
  \bibinfo{author}{Akata, Z.}, \bibinfo{author}{Stumpf, S.},
  \bibinfo{author}{Kieseberg, P.}, \& \bibinfo{author}{Holzinger, A.}
  (\bibinfo{year}{2018}).
\newblock \bibinfo{title}{Explainable {AI}: the new 42?}
\newblock In {\it \bibinfo{booktitle}{Machine Learning and Knowledge
  Extraction: Second IFIP TC 5, TC 8/WG 8.4, 8.9, TC 12/WG 12.9 International
  Cross-Domain Conference, CD-MAKE 2018, Hamburg, Germany, August 27--30, 2018,
  Proceedings 2}\/} (pp. \bibinfo{pages}{295--303}).
\newblock \bibinfo{organization}{Springer}.
\bibitem[{Henderson et~al.(2016)Henderson, Child, Moore, Moore \&
  Kaczynski}]{henderson2016influence}
\bibinfo{author}{Henderson, H.}, \bibinfo{author}{Child, S.},
  \bibinfo{author}{Moore, S.}, \bibinfo{author}{Moore, J.~B.}, \&
  \bibinfo{author}{Kaczynski, A.~T.} (\bibinfo{year}{2016}).
\newblock \bibinfo{title}{The influence of neighborhood aesthetics, safety, and
  social cohesion on perceived stress in disadvantaged communities}.
\newblock {\it \bibinfo{journal}{American Journal of Community Psychology}\/},
  {\it \bibinfo{volume}{58}\/}, \bibinfo{pages}{80--88}.
\bibitem[{Jain et~al.(2023)Jain, Mohankumar, Wan, Ganguly, Wilson \&
  Anderson}]{jain2023training}
\bibinfo{author}{Jain, M.}, \bibinfo{author}{Mohankumar, N.~M.},
  \bibinfo{author}{Wan, H.}, \bibinfo{author}{Ganguly, S.},
  \bibinfo{author}{Wilson, K.~D.}, \& \bibinfo{author}{Anderson, D.~M.}
  (\bibinfo{year}{2023}).
\newblock \bibinfo{title}{Training machine learning models to characterize
  temporal evolution of disadvantaged communities}.
\newblock \bibinfo{note}{ArXiv preprint arXiv:2303.03677}.
\bibitem[{Jolliffe(1982)}]{jolliffe1982note}
\bibinfo{author}{Jolliffe, I.~T.} (\bibinfo{year}{1982}).
\newblock \bibinfo{title}{A note on the use of principal components in
  regression}.
\newblock {\it \bibinfo{journal}{Journal of the Royal Statistical Society
  Series C: Applied Statistics}\/},  {\it \bibinfo{volume}{31}\/},
  \bibinfo{pages}{300--303}.
\bibitem[{Jolliffe \& Cadima(2016)}]{jolliffe2016principal}
\bibinfo{author}{Jolliffe, I.~T.}, \& \bibinfo{author}{Cadima, J.}
  (\bibinfo{year}{2016}).
\newblock \bibinfo{title}{Principal component analysis: a review and recent
  developments}.
\newblock {\it \bibinfo{journal}{Philosophical Transactions of the Royal
  Society A: Mathematical, Physical and Engineering Sciences}\/},  {\it
  \bibinfo{volume}{374}\/}, \bibinfo{pages}{20150202}.
\bibitem[{Kawano et~al.(2018)Kawano, Fujisawa, Takada \&
  Shiroishi}]{kawano2018sparse}
\bibinfo{author}{Kawano, S.}, \bibinfo{author}{Fujisawa, H.},
  \bibinfo{author}{Takada, T.}, \& \bibinfo{author}{Shiroishi, T.}
  (\bibinfo{year}{2018}).
\newblock \bibinfo{title}{Sparse principal component regression for generalized
  linear models}.
\newblock {\it \bibinfo{journal}{Computational Statistics \& Data Analysis}\/},
   {\it \bibinfo{volume}{124}\/}, \bibinfo{pages}{180--196}.
\bibitem[{Leyshon \& Thrift(1994)}]{leyshon1994access}
\bibinfo{author}{Leyshon, A.}, \& \bibinfo{author}{Thrift, N.}
  (\bibinfo{year}{1994}).
\newblock \bibinfo{title}{Access to financial services and financial
  infrastructure withdrawal: problems and policies}.
\newblock {\it \bibinfo{journal}{Area}\/},  (pp. \bibinfo{pages}{268--275}).
\bibitem[{Lipton(2018)}]{lipton2018mythos}
\bibinfo{author}{Lipton, Z.~C.} (\bibinfo{year}{2018}).
\newblock \bibinfo{title}{The mythos of model interpretability: In machine
  learning, the concept of interpretability is both important and slippery.}
\newblock {\it \bibinfo{journal}{Queue}\/},  {\it \bibinfo{volume}{16}\/},
  \bibinfo{pages}{31--57}.
\bibitem[{Mandarano \& Meenar(2017)}]{mandarano2017equitable}
\bibinfo{author}{Mandarano, L.}, \& \bibinfo{author}{Meenar, M.}
  (\bibinfo{year}{2017}).
\newblock \bibinfo{title}{Equitable distribution of green stormwater
  infrastructure: A capacity-based framework for implementation in
  disadvantaged communities}.
\newblock {\it \bibinfo{journal}{Local Environment}\/},  {\it
  \bibinfo{volume}{22}\/}, \bibinfo{pages}{1338--1357}.
\bibitem[{Murali \& Oyebode(2004)}]{murali2004poverty}
\bibinfo{author}{Murali, V.}, \& \bibinfo{author}{Oyebode, F.}
  (\bibinfo{year}{2004}).
\newblock \bibinfo{title}{Poverty, social inequality and mental health}.
\newblock {\it \bibinfo{journal}{Advances in Psychiatric Treatment}\/},  {\it
  \bibinfo{volume}{10}\/}, \bibinfo{pages}{216--224}.
\bibitem[{Osberg(1985)}]{osberg1985measurement}
\bibinfo{author}{Osberg, L.} (\bibinfo{year}{1985}).
\newblock \bibinfo{title}{The measurement of economic well-being}.
\newblock \bibinfo{publisher}{University of Toronto Press}.
\bibitem[{Papanicolaou(2009)}]{papanicolaou2009taylor}
\bibinfo{author}{Papanicolaou, A.} (\bibinfo{year}{2009}).
\newblock \bibinfo{title}{Taylor approximation and the delta method}.
\bibitem[{Qian \& Jaller(2020)}]{qian2020bikesharing}
\bibinfo{author}{Qian, X.}, \& \bibinfo{author}{Jaller, M.}
  (\bibinfo{year}{2020}).
\newblock \bibinfo{title}{Bikesharing, equity, and disadvantaged communities: A
  case study in chicago}.
\newblock {\it \bibinfo{journal}{Transportation Research Part A: Policy and
  Practice}\/},  {\it \bibinfo{volume}{140}\/}, \bibinfo{pages}{354--371}.
\bibitem[{Ray(1998)}]{ray1998development}
\bibinfo{author}{Ray, D.} (\bibinfo{year}{1998}).
\newblock {\it \bibinfo{title}{Development economics}\/}.
\newblock \bibinfo{publisher}{Princeton University Press}.
\bibitem[{Tibshirani(1996)}]{tibshirani1996regression}
\bibinfo{author}{Tibshirani, R.} (\bibinfo{year}{1996}).
\newblock \bibinfo{title}{Regression shrinkage and selection via the lasso}.
\newblock {\it \bibinfo{journal}{Journal of the Royal Statistical Society
  Series B: Statistical Methodology}\/},  {\it \bibinfo{volume}{58}\/},
  \bibinfo{pages}{267--288}.
\bibitem[{Vidal(1995)}]{vidal1995reintegrating}
\bibinfo{author}{Vidal, A.~C.} (\bibinfo{year}{1995}).
\newblock \bibinfo{title}{Reintegrating disadvantaged communities into the
  fabric of urban life: The role of community development}.
\newblock {\it \bibinfo{journal}{Housing Policy Debate}\/},  {\it
  \bibinfo{volume}{6}\/}, \bibinfo{pages}{169--230}.
\bibitem[{Wiesel et~al.(2018)Wiesel, Liu \& Buckle}]{wiesel2018locational}
\bibinfo{author}{Wiesel, I.}, \bibinfo{author}{Liu, F.}, \&
  \bibinfo{author}{Buckle, C.} (\bibinfo{year}{2018}).
\newblock \bibinfo{title}{Locational disadvantage and the spatial distribution
  of government expenditure on urban infrastructure and services in
  metropolitan {S}ydney (1988--2015)}.
\newblock {\it \bibinfo{journal}{Geographical Research}\/},  {\it
  \bibinfo{volume}{56}\/}, \bibinfo{pages}{285--297}.

\end{thebibliography}

\clearpage

\end{document}